\documentstyle[preprint,aps,epsfig,rotating]{revtex}

\begin{document}
\draft
\preprint{}
\title{Two dimensional nature of superconductivity in intercalated layered systems 
Li$_{x}$HfNCl and Li$_{x}$ZrNCl:  
muon spin relaxation and magnetization measurements}
\author{
T.~Ito\cite{byline}, Y.~Fudamoto, A.~Fukaya, I.M.~Gat-Malureanu, 
M.I.~Larkin, P.L.~Russo, A.~Savici, and Y.J.~Uemura\cite{byline2}
}
\address{
Department of Physics, Columbia University, 538W 120th St., New York, NY 10027
}
\author{
K.~Groves, and R.~Breslow
}
\address{
Department of Chemistry, Columbia University, 3000 Broadway, New York, NY 10027
}
\author{
K.~Hotehama and S.~Yamanaka
}
\address{
Department of Applied Chemistry, Graduate School of Engineering, 
Hiroshima University, Higashi-Hiroshima 
739-8527, Japan
}
\author{
P.~Kyriakou, M.~Rovers, and G.M.~Luke
}
\address{
Department of Physics and Astronomy, McMaster Univ., Hamilton, ON L8S 4M1, 
Canada
}
\author{
K.M.~Kojima
}
\address{
Department of Superconductivity, University of Tokyo, Bunkyo-ku, 
Tokyo 113-8656, Japan
}

\date{\today}
\maketitle

\begin{abstract}
We report muon spin relaxation ($\mu$SR) and magnetization
measurements, together with synthesis and characterization, 
of the Li-intercalated layered superconductors
Li$_{x}$HfNCl and Li$_{x}$ZrNCl with/without   
co-intercalation of THF (tetrahydrofuran) or PC (propylene carbonate).   
The 3-dimensional (3-d) superfluid density
$n_{s}/m^{*}$ (superconducting carrier density / effective mass),
as well as the two dimensional superfluid density
$n_{s2d}/m^{*}_{ab}$ (2-dimensional (2-d) area density of superconducting
carriers / ab-plane effective mass), have been derived from the 
$\mu$SR results of the 
magnetic-field penetration depth $\lambda_{ab}$ observed with  
external magnetic field applied perpendicular to the 2-d 
honeycomb layer of HfN / ZrN.  
In a plot of $T_{c}$ versus $n_{s2d}/m^{*}_{ab}$, most of
the results lie close to the linear relationship
found for underdoped high-$T_{c}$ cuprate (HTSC) 
and layered organic BEDT superconductors.
In Li$_{x}$ZrNCl without THF intercalation, 
the superfluid density and $T_{c}$ for
$x$ = 0.17 and 0.4 do not show much difference, reminiscent
of $\mu$SR results for some overdoped HTSC systems.    
Together with the absence of dependence of $T_{c}$ on
average interlayer distance among ZrN / HfN layers, 
these results suggest that 
the 2-d
superfluid density $n_{s2d}/m^{*}_{ab}$ is a 
dominant determining factor for $T_{c}$ in the 
intercalated nitride-chloride systems.
We also report 
$\mu$SR and magnetization results on depinning of flux vortices,
and the magnetization results for the 
upper critical field $H_{c2}$ and the 
penetration depth $\lambda$.  Reasonable agreement was obtained
between $\mu$SR and magnetization estimates of $\lambda$. 
We discuss the two dimensional nature of superconductivity in 
the nitride-chloride systems based on these results. 
  
\end{abstract}

\narrowtext

\section{INTRODUCTION}
\label{sec:level1}

Layered superconductors, such as high-$T_{c}$ cuprates (HTSC)
or organic BEDT systems, have been a subject of extensive
research effort for decades.  
These systems show rich novel phenomena, including 
superconducting fluctuations, pancake vortex, 
complicated vortex phase diagrams, and 
interlayer Josephson effects. High-$T_{\rm c}$ 
cuprate superconductors have been investigated extensively 
as prototypical layered superconductors. 
The cuprates have a merit that their carrier concentration can be 
controlled by chemical substitutions and/or oxygen contents. 
On the other hand, it has not been easy to control the interlayer distance
in cuprates.
In general, superconductivity of these layered systems is
deeply related to in-plane features as well as interlayer
couplings.  For overall understanding of 
superconductivity in these systems, it would be essential
to elucidate interplay between in-plane and interplane 
properties.  Despite extensive research effort, however, 
detailed roles
of dimensionality are yet to be clarified in these systems.

Recently, superconductivity was discovered 
in ZrNCl and HfNCl intercalated with
alkali atoms (Li, Na, K)  
\cite{discovery_Zr,discovery_Hf}.  Systems based on ZrNCl have  
superconducting transition temperatures $T_{c} \leq 
15$ K, while those based on HfNCl have $T_{c} \leq 25.5$ K.
These systems can be co-intercalated with organic molecules,
such as THF (tetrahydrofuran) or PC (propyrene carbonate).
The parent compounds ZrNCl and HfNCl are insulators 
which have a layered structure as shown in Fig. 1(a). 
Zr(Hf)-N honeycomb double layers are sandwiched by Cl layers, 
and a composite Cl-(ZrN)-(ZrN)-Cl layers form a stacking unit. 
Adjacent stacking units are bonded by weak van der Waals force. 
Alkali metal atoms and polar organic molecules such as THF
or PC can be co-intercalated into the 
van der Waals gap of the parent compounds as shown schematically 
in Fig. 1(b). Intercalated alkali metal atoms are supposed 
to release electrons into Zr(Hf)-N double layers, 
which makes the system metallic and superconducting. 
On the other hand, 
intercalated organic molecules expand interlayer distance 
without changing Zr(Hf)-N honeycomb double layers. 
So we can control two separate parameters,
carrier concentration and the stacking unit distance, in a single series of 
nitride chlorides with the common superconducting slab.
This unique feature could 
allow studies of layered superconductors from a new angle. 

In this paper, we will present synthesis and characterization of a series of 
intercalated ZrNCl and HfNCl samples, 
together with studies of
their superconducting 
properties using muon spin relaxation 
($\mu$SR), magnetization, and resistivity 
measurements. A part of this work was presented in a conference 
\cite{ssc20}, where we reported $\mu$SR results in
HfNCl-Li$_{0.5}$-THF$_{0.3}$, showed that $T_{c}$ and the 2-d superfluid 
density
in this system follow the correlations found in cuprates and BEDT systems,
and discussed that this feature likely comes from departure from BCS 
condensation,
which can be understood in terms of crossover from Bose-Einstein to BCS
condensation.  Subsequently, Tou {\it et al.\/} \cite{touprl} 
reported an NMR Knight shift study
which inferred a rather small density of states at the Fermi level in 
HfNCl-Li-THF, and discussed difficulty in explaining the high transition 
temperature $T_{c}$ in terms of the conventional BCS theory.  
Tou {\it et al.\/} \cite{touprb}
also reported a rather high upper 
critical field $H_{c2}(T\rightarrow 0) \sim 100$ kG,
for the field applied perpendicular to the conducting planes, 
from magnetization and NMR measurements.    
 
Extensive $\mu$SR measurements of the 
magnetic field penetration depth $\lambda$ have been 
performed to date in 
various superconducting systems, 
such as high-$T_{c}$ cuprates 
(HTSC) \cite{ssc6,ssc7,ssc8,ssc12,ssc13,ssc14,ssc15,ssc25}, 
fullerides \cite{ssc17,ssc18},  
and two-dimensional (2-d) organic BEDT systems \cite{ssc19}. 
Universal nearly linear correlations have
been found between 
$T_{c}$ and the muon spin relaxation rate 
$\sigma(T \to 0) \propto 1/\lambda^{2} \propto n_{s}/m^{*}$
(superconducting carrier density / effective mass)  
in the underdoped
region of many HTSC systems and in some other exotic 
superconductors \cite{ssc6,ssc7}
Such correlations are seen also 
in HTSC superconductors having extra perturbations, 
such as overdoping \cite{ssc8,ssc15,ssc23}, 
(Cu,Zn) substitutions \cite{ssc9}, or spontaneous formation of nano-scale
regions with static stripe spin correlations \cite{ssc10,ssc11}. 
In all these systems, $T_{c}$ follows the 
correlations with superfluid density found for less perturbed 
standard HTSC systems. These results indicate that
the superfluid density is a determining factor for $T_{c}$
in the cuprates \cite{sscreview}.

In general, a strong dependence of $T_{c}$ on the carrier density
is not expected in conventional BCS theory\cite{ssc45} where $T_{c}$ is
determined by the mediating boson (phonon) energy scale and the density of 
states of carriers at the Fermi level which govern 
the charge-boson (electron-phonon) coupling.  For 2-d 
metals, the density of states does not depend on the carrier density
in the simplest case of non-interacting fermion gas.  
Therefore, the BCS theory has a fundamental difficulty in explaining the 
observed correlations.
In contrast, 
an explicit dependence of $T_{c}$ can be expected for the 
condensation temperature $T_{B}$ in Bose-Einstein (BE) condensation 
of a simple Bose gas,
as well as for the Kosterlitz-Thouless transition temperature $T_{KT}$ for a 
2-d superfluid \cite{ssc40}.  
In BE and KT transitions, the transition temperature
is determined simply by the number density and the mass, since the
condensation is decoupled from the formation of condensing bosons in these
two cases.  The universality of the $T_{c}$ vs $n_{s}/m^{*}$ relationship 
observed
beyond the difference of systems, such as cuprates, fullerides, 
orangics, etc., may be related to this feature. 
Pictures proposed to explain
the correlations between $T_{c}$ and the superfluid density in the 
cuprates include  
crossover from Bose-Einstein to BCS 
condensation \cite{ssc47,ssc48,ssc42,ssc49,ssc50}
and phase fluctuations \cite{ssc51}.  In the present work, we extend our study  
to intercalated nitride-chloride systems,
seeking further insights into such phenomenology.

We have also determined the upper critical field $H_{c2}$ 
of nitride-chloride systems from magnetization and resistivity measurements. 
Although it is not easy to determine   
$H_{c2}$ in layered superconductors due to the strong superconducting 
fluctuations, Hao {\it et al.} \cite{hao} developed an approach to overcome
such difficulty using a model for 
reversible diamagnetic magnetization of type-II 
superconductors which have high $\kappa$ values.
Here, $\kappa$ is the Ginzburg-Landau 
parameter defined as the ratio of the penetration depth $\lambda$ 
to the coherence length $\xi$. 
In this model, one calculates the free energy including 
the supercurrent kinetic energy, the magnetic-field energy,
as well as the kinetic-energy and condensation-energy terms arising 
from suppression of the order parameter in the vortex core. 
This method has been successfully applied to various 
high-$T_{c}$ cuprate superconductors. We will apply this model to 
superconducting Hf(Zr)NCl in the temperature and field region 
where the effect of superconducting fluctuations is negligible.

\section{SYNTHESIS AND CHARACTERIZATION}
\label{sec:level2}

The parent compounds HfNCl and ZrNCl were synthesized by the reaction
of Zr or Hf powder with vaporized NH$_{4}$Cl at 
600\ $^{\circ}$C for 30 minutes 
in N$_{2}$ flow \cite{prereaction}. The resulting powder was 
sealed in a quartz ampule and was purified by a chemical 
vapor transport method with temperature gradient 
\cite{chemical_transport}. 
For the purification, an end of the ampule with pre-reacted 
powder was kept at 800\ $^{\circ}$C and the other end, 
where purified powder is collected, at 900\ $^{\circ}$C 
for three days. No impurity phase was detected in the purified 
powder from x-ray diffraction. 

Since a Li-intercalated sample is sensitive to air, 
intercalation was performed in a glove box. 
We used three intercalation methods \cite{intercalation} to prepare 
a variety of samples: 
(i) Li-intercalated ZrNCl samples were prepared 
by soaking parent ZrNCl powder in three kinds of butyllithium
solution. We used 2.0 $M$ ($mol/l$) $n$-butyllithium solution in 
cyclohexane, 1.3 $M$ $sec$-butyllithium solution in 
cyclohexane, and 1.7$M$ $tert$-butyllithium solution in 
pentane for this purpose. 
In this order, reducing ability becomes stronger and therefore 
higher concentration of Li atoms can be intercalated into 
the parent compound. 
We used solution which contained butyllithium corresponding to more than 
2, 5, and 5 equivalents of ZrNCl for $n$-, $sec$-, and 
$tert$-butyllithium, to avoid the dilution of butyllithium 
in the intercalation process. 
(ii) We performed co-intercalation of Li and organic molecules into ZrNCl 
by soaking Li intercalated ZrNCl powder (prepared by 
the method (i) using $n$-Butyllithium solution) in 
enough THF or PC. 
(iii) Li- and THF-co-intercalated HfNCl samples were prepared 
by soaking HfNCl powder in various concentrations 
(2.5-100 $mM$) of lithium naphthalene (Li-naph) solution in THF. 
Since phase separation was often observed in the samples 
prepared by the method (iii), we selected single phase samples 
for the measurements in this study via characterization 
from $T_{c}$ as well as 
c-axis lattice constant. The Li- and THF-co-intercalated HfNCl 
sample with the largest c-axis lattice constant according to 
the method (iii) was prepared 
at Hiroshima University, while all the other samples at Columbia University. 
Table 1 shows a list of these samples. 

The chemical composition was determined by inductively coupled 
plasma atomic emission spectroscopy (ICP-AEM) and CHN elemental 
analysis. 
The powder samples were pressed into pellets under uniaxial 
stress and sealed in cells made of Kapton film and epoxy glue 
for x-ray measurements. 
The x-ray rocking curve of $(00l)$ peaks for cleaved surface 
(inside of a pellet) 
shows a peak with half width at half maximum of $\sim$ 8 degrees,
which indicates that the bulk of the samples have well-aligned 
preferred orientation and are suitable for studies of 
anisotropic properties. Essentially similar rocking curves were 
also observed for as-prepared surfaces of pellets.  

The superconducting transition temperature 
of the samples was determined from measurements of magnetic
susceptibility $\chi$. 
In Fig. 2, we show the results of $\chi$ in intercalated 
ZrNCl specimens which were examined by $\mu$SR measurements.  
The values of  
$\chi = M/H$ are of the order for ideal perfect diamagnetism 
$-3/8\pi$ for spherical samples and
$-1/4\pi$ for long cylindrical samples 
with the field parallel to the cylinder axis.  
The determination of the absolute values of $\chi$, however, can be
affected by such factors as non-spherical sample shape, sample morphology and 
residual field in a SQUID magnetometer.  These factors might have 
caused deviation of some of the zero-field cooled shielding
values of $\chi$ from $-1/4\pi$.

In Table 1, we summarize the composition, synthesis method, 
distance between adjacent stacking units  
(1/3 of the c-axis lattice constant $c_{0}$, see Fig. 1), and 
the superconducting transition temperature $T_{c}$ 
for these samples. 
The stacking unit distance is 
9.4 \AA\ for the Li-intercalated samples prepared by
the method (i) without co-intercalation of organic molecules.
This value is almost the same 
as that for unintercalated parent compound (9.3 \AA) without Li. 
With increasing Li concentration $x$ from 0.17 to 0.6 (in the samples 
without co-intercalation of organic molecules), 
$T_{c}$ decreases from 14.2 K to 11.7 K.
Figure 3(a) shows the $x$ 
dependence of $T_{c}$ in Li$_{x}$ZrNCl and Li$_{x}$HfNCl
samples, with/without co-intercalation, obtained in the 
present study.
In Fig. 3 and Table 1, we find that:
(1) $T_{c}$ shows a slow reduction with increasing $x$;
(2) for close $x$ values, $T_{c}$ does not depend 
much on the stacking-unit 
distance $c_{o}$/3 (see Fig. 3(b)).
These results are qualitatively
consistent with the reported results  
in the Li, K, and Na doped ZrNCl
samples with/without co-intercalation of organic molecules \cite{kawaji}.  
Decrease of $T_{c}$ with increasing charge doping is 
reminiscent of the case of overdoped high-$T_{c}$ cuprate superconductors. 

To the best of our knowledge, this is the first report of success 
in Li-THF intercalation into ZrNCl by $sec$- and $tert$-butyllithium. 
The stacking unit thickness is 13.3 or 18.7 \AA\ for the 
methods (ii) and (iii). As has been reported, 
$T_{c}$ is almost unaffected by the expansion of 
the stacking-unit distance from $\sim$ 9.4 \AA\ to $\sim$ 13.3 \AA\ 
for the samples with Li content $x \sim 0.17$ (see Fig.\ 2 (b)). 
Systems based on HfNCl has $T_{c}$ = 25.5 K, 
nearly a factor of two 
higher than that for intercalated ZrNCl. 

\section{$\mu$SR: EXPERIMENTAL}
\label{sec:level3}

Our $\mu$SR experiments were 
performed at TRIUMF, the Canadian National Accelerator Laboratory 
located in Vancouver, Canada, 
which provided a high intensity and polarized beam of positive muons. 
Each pressed pellet sample,
with the c-axis aligned, was sealed in a sample cell 
which has a Kapton window 
and mounted in a He gas-flow cryostat with the c-axis parallel to the direction of muon beam.
Transverse external field (TF) 
was applied parallel to the beam direction, while
muons are injected with their initial spin polarization perpendicular to the
field/beam direction.  Low-momuntum (surface) muons with the incident
momentum of 29.8 MeV/c were implanted in the pellet specimens. The
average stopping depth, 100-200 mg/cm$^{2}$, assured that the majority of
muons are stopped within the specimen, after going through Kapton windows of the cryostat and
the sample cell.
  
Plastic sintillation counters were used to 
detect the arrival of a positive muon and its decay into a positron,
and the decay-event histogram was obtained, as a function of 
muon residence time $t$ which corresponds to the time difference
of the muon arrival and positron decay signals.
The time evolution of muon spin direction/polarization was
obtained from the angular asymmetry of positron histograms, after 
correction for 
the exponential decay $\exp(-t/\tau_{\mu})$, where $\tau_{\mu}$ = 2.2
$\mu$s is the mean lifetime of a positive muon.
Details of $\mu$SR technique can be found, for example, in refs 
\cite{ssc1,ssc2,ssc3}.

The asymmetry time spectra $A(t)$ were fit to a functional 
form;
$$A(t)=A(0)\exp(-\sigma^{2}t^{2}/2)\times cos(\omega t + \phi),$$
where $A(0)$ is the initial decay asymmetry at $t$ = 0.
The muon spin precesses at the frequencies 
$\omega = \gamma_{\mu}H_{ext}$, where $\gamma_{\mu}$ 
is the gyromagnetic ratio of a muon 
($2\pi \times 13.554$ MHz/kG) and $H_{ext}$ 
denotes the transverse external magnetic field.
As shown in Fig. 4, for an example of 
Li$_{0.17}$ZrNCl, this oscillation exhibits faster damping in the 
superconducting state due to 
inhomogeneous distribution of internal magnetic
fields in the flux vortex structure.
In pressed pellet samples of random or oriented
powder, this relaxation can usually be approximated 
by a Gaussian decay which defines the muon spin 
relaxation rate $\sigma$. 

For systems except for ZrNCl-Li-THF, the $\mu$SR results were
analysed by assuming a single-component signal, which 
shows a reasonble agreement to the data as in Fig. 4(b).
In ZrNCl co-intercalated with Li and THF, the relative value of 
the shielding susceptibility was significantly lower than
those of other samples, as shown in Fig. 2.  Although it is 
not clear, this reduced susceptibility could possibly 
imply a finite fraction of superconducting volume.
As a precautionary measure, by fitting selected low-temperature signals in field-cooled
and zero-field cooled procedures to an asymmetry function having
two-component signals, we estimated an upper-limit
of the relaxation rate for the superconducting fraction.
This upper-limit is shown (in Figs. 6 and 7) by the error-bar placed to the 
right side of the main point for $\sigma$ which was obtained
for a single-component asymmetry function.   
 
\section{$\mu$SR: SPECTRA AND RELAXATION RATE}
\label{sec:level4}

Figure 4 shows time spectra of muon decay asymmetry for 
a representative sample 
above and below $T_{c}$. 
In the normal state above $T_{c}$, the oscillation shows 
a small relaxation due to nuclear dipole fields.
We denote this relaxation rate as $\sigma_{n}$. 
Below $T_{c}$, the relaxation becomes faster due to 
additionaal field distribution from the flux vortex lattice.  
For each specimen, the zero-field $\mu$SR spectra obtained 
above and well below $T_{\rm c}$ did not show any difference.
This assures that the temperature dependence of the  
relaxation rate observed in TF is due to 
superconductivity alone, and also implies that 
there is no detectable effect of time-reversal symmetry 
breaking, contrary to the case of UPt$_3$ 
\cite{UPt3} and Sr$_2$RuO$_4$ \cite{Sr2RuO4}. 

The effect of the superconducting vortex lattice 
can be obtained by subtracting this normal-state background $\sigma_{n}$ from 
the observed relaxation rate $\sigma_{ob}$.
Since the nuclear dipolar broadening and superconducting 
broadening of the internal fields do not add coherently, here we adopt 
quadratic subtraction
to obtain the relaxation rate $\sigma$ due to superconductivity as:
$$\sigma = \sqrt{\sigma_{ob}^2-\sigma_{n}^2}\ \ \ 
for\ \ \ (\sigma_{ob} \ge \sigma_{n})$$ 
and 
$$\sigma = -\sqrt{\sigma_{n}^2-\sigma_{ob}^2}\ \ \ 
for\ \ \ (\sigma_{ob} < \sigma_{n}).$$ 
Note that this procedure 
makes the error bar rather large around $\sigma$ = 0. 

Figure 5(a) shows the temperature dependence of the relaxation rate $\sigma$ 
for the samples co-intercalated with organic molecules
having expanded 
interlayer distance. With decreasing temperature, 
the relaxation rate begins to increase below the superconducting 
transition temperature $T_{c}$. At the flux-pinning 
temperature $T_{p}$, the zero-field-cooling (ZFC) curve 
begins to deviate from the field-cooling (FC) curve.  
In the ZFC procedures, flux vortices are required to 
enter the specimen from its edge and move a large
distance before reaching their equilibrium position. 
Below the pinning temperature $T_{p}$, this long-distance 
flux motion could be prevented by the flux pinning, resulting
in an highly inequilibrium flux lattice and more inhomogeneous field distribution
at muon sites. This behavior has been observed in earlier $\mu$SR
studies of HTSC \cite{wuprbbi2212}, BEDT \cite{ssc19} and some other systems.
In both systems shown in Fig. 5(a), we find that $T_{p}$ is much lower
than $T_{c}$, which is a characteristic feature for
highly 2-d superconductors.
The $\mu$SR results of $T_{p}$ for the present 
systems agree well with those from magnetization 
measurements discussed in Section VII. 

We performed FC measurements using a wide range of 
external transverse magnetic fields $H_{ext}$, and found 
no significant dependence of $\sigma$ on $H_{ext}$  
from 40 G to 1000 G, as shown in Fig. 5.
In TF-$\mu$SR measurements in highly 2-d
superconductors, such as Bi2212 or (BEDT-TTF)$_{2}$Cu(NCS)$_{2}$,
application of a high external magnetic field transforms 3-d flux vortex
structure into 2-d pancake vortices, since higher field implies stronger
coupling of flux vortices within a given plane and higher chance for 
the vortex location in each plane to be determined by random defect position 
on each plane \cite{3Dvortex}.   
The absence of field dependence in our measurements implies that 
corrections for the 2-d vortex effect is not necessary in the present
study.  This situation is expected for our c-axis aligned powder specimens. 
The relaxation rate $\sigma(T)$ shows a tendency of saturation at 
low temepratures in all of the 
measured samples of nitride-chloride systems 
in the present study. This behavior is characteristic for s-wave 
superconductors. However, experiments using high-quality single crystals are 
required for a conclusive determination of the superconducting 
pairing symmetry.
In the case of HTSC cuprates, d-wave pairing was established only 
after $\mu$SR results on high-quality crystals of YBa$_{2}$Cu$_{3}$O$_{y}$
became available \cite{ssc4}.

Figure 5(b) shows the temperature dependence of $\sigma$ 
for the samples without organic co-intercalant.
The dependence of $\sigma$ on temperature $T$ and field
$H_{ext}$ of the field-cooling results was
essentially similar to that for specimens with 
co-intercalation in Fig. 5(a).  In these systems, we 
determined the pinning temperature $T_{p}$ by 
magnetization measurements instead of by $\mu$SR, 
and show the results in section VII.

\section{$\mu$SR: COMPARISON WITH OTHER SYSTEMS AND SUPERFLUID ENERGY SCALES}
\label{sec:level5}

The $\mu$SR relaxation rate due to the penetration depth is 
related to the superconducting carrier density $n_s$, 
effective mass $m^*$, the coherence length $\xi$ and 
the mean free path $l$ as 
$$
\sigma \propto \lambda^{-2} = 
\frac{4\pi e^2}{c^2} \times \frac{n_{s}}{m^{*}} 
\times \frac{1}{1+\xi/l}.
$$
The proportionality to $n_{s}/m^{*}$ comes from the fact that
this effect is caused by the superconducting screening current,
and consequently reflecting the current density in a similar
way to the normal state conductivity of a metal which is proportional
to the carrier density divided by the effective mass. 
As will be shown later, $\xi$ is estimated to be 80-90 \AA\ in 
the present nitride-chloride systems. 
The mean free path cannot be determined at the moment, 
since a high-quality single crystal is not yet available. 
In this situation, we proceed with the following arguments by 
assuming that the system falls in the clean limit 
($\xi << l$). Clean limit has been confirmed
in many other strongly
type-II superconductors, such as the cuprates and BEDT systems.

In highly anisotropic 2-d superconductors, 
the penetration depth measured with the external field
parallel and perpenducular to the conducting
plane could be very different.  For the geomentry
with $H_{ext}$ perpendicular to the conducting plane,
related to the in-plane penetration depth $\lambda_{ab}$
as in the present study, the superconducting screening
current flows within the plane, resulting in 
the more effective partial screening of $H_{ext}$ and the
shorter $\lambda$ compared to the case with 
$H_{ext}$ parallel to the planes.  In the present work,
our specimen has a highly-oriented c-axis direction within 
+/- 8 degrees, and we regard our specimen as equivalent to 
single crystal specimens in terms of anisotropy.
A theory/simulation work \cite{ssc5} shows that for un-oriented
ceramic specimens of highly 2-d superconductors, 
value of $\sigma$ should be reduced by a factor of 1/1.4
from the value for single crystals observed with 
$H_{ext}$ applied perpenducular to the conducting planes.
      
In Fig. 6, in a plot of $T_{c}$ versus 
the low-temperature relaxation rate 
$\sigma(T \rightarrow 0) \propto n_{s}/m^{*}$, we 
compare the results of the present nitride-chloride systems 
with those from cuprate and organic BEDT superconductors.
The point for the BEDT system was obtained in $\mu$SR
measurements using single crystal specimens \cite{ssc19}.
The shaded area denoted as cuprates represents the
universal linear correlations found for  
un-oriented ceramic specimens of 
underdoped YBCO systems: we multiplied the relaxation 
rate in these YBCO by a factor of 1.4 to account for the difference between
single crystal and un-oriented ceramic specimens.  
The data points lie in possibly two different groups having different
slopes in the $T_{c}$/$\sigma$ relation.
The first group with a higher slope includes the 
present nitride-chloride systems with co-intercalation
of organic molecules and the BEDT system, all of which having
highly 2-d character as demonstrated by their
depinning temperature $T_{p}$ being nearly a 1/3 to 1/2 of $T_{c}$.
The second group includes YBCO cuprates and nitride-chloride
systems without organic co-intercalation, all of which have
more 3 dimensional (3-d) character in the flux pinning property
with $T_{p}$ closer to $T_{c}$.
The irreversibility and depinning behavior in Li$_{x}$ZrNCl
without organic co-intercalation was studied not by $\mu$SR
but by magnetization measurements as described in 
section VII.

The relaxation rate observed by $\mu$SR is determined by 
the 3-d superfluid density $n_{s}/m^{*}$,
as this is a phenomenon caused by the screening supercurrent
density in bulk specimens.  With the knowledge of 
interlayer spacing $c_{int}$, one can convert 3-d density 
$n_{s}/m^{*}$ into 2-d density on each conducting 
plane as $n_{s2d}/m^{*} = n_{s}/m^{*} \times c_{int}$.
For systems having double-layer conducting planes, such as 
the present nitride chlorides or some family of the cuprates,
the average interlayer spacing depends on whether or not
the double layer is regarded as a single conducting unit or
two.  In our previous reports for cuprates \cite{ssc42}, we treated 
the double layer as two single layers.  We shall follow
this approach here, and define the $c_{int}$ to be a half of
the stacking unit distance as $c_{int} = c_{0}/6$.

In Fig. 7, we show a plot of $T_{c}$ versus the 
2-d superefluid density $n_{s2d}/m^{*}$ represented by 
the value $\sigma(T\rightarrow 0) \times c_{int}$.
We include a point obtained in c-axis oriented ceramic
specimen of YBa$_{2}$Cu$_{3}$O$_{7}$ \cite{YBC_oriented}
(without multiplying a factor 1.4 to $\sigma$ since this
specimen had an almost perfect alignment of c-axis direction).
We find that most of the data points share a unique slope  
in Fig. 7.  
This result suggests two features:
(1) within the nitride-chloride systems, 2-d superfluid density
$n_{s2d}/m^{*}$ is a determining factor for $T_{c}$;
and (2) the 2-d superfluid density may 
even be a fundamental determining factor for $T_{c}$  
among different superconducting systems.
However, the second conclusion (2) must be taken 
with caution, because, this analysis depends on 
our treatment regarding single versus double layers,
and also because recent data on Tl2201 \cite{ssc8}.
and Bi2201 \cite{tbp} cuprates, having
very large interlayer distance $c_{int} > $12 \AA, show universal
behavior with the results from YBCO ($c_{int} \sim$ 6 \AA)
only in a 3-d plot like Fig. 6 but not in a 2-d plot like Fig. 7 \cite{ssc42}.
In contrast, the conclusion (1) is more robust, since all 
the nitride-chloride
systems have double conducting layers, and since the predominant
2-d character is consistent with the absence of dependence of $T_{c}$ on 
interlayer spacing in nitride-chlorides shown in Fig. 3(b).

Since the Fermi energy of a 2-d metal is proportional to the 
2-d carrier density $n_{2d}$ divided by the in-plane effective
mass $m^{*}$, as $T_{F} = (\hbar^{2}\pi)n_{2d}/m^{*}$, the horizontal axis of 
Fig. 7 can be converted into an energy scale representing 
superconducting condensate.  This conversion from 
penetration depth to the superfluid energy scale was first
attempted by Uemura {\it et al.\/} \cite{ssc7} in 1991 and 
later followed by other researchers, including
Emery and Kivelson \cite{ssc51}.   

In order to do such a conversion, one needs to obtain 
absolute values of the penetration depth $\lambda$ from the 
relaxation rate $\sigma$.
The numerical factor in this $\sigma$ to $\lambda$ conversion  
in $\sigma \propto \lambda^{-2}$
depends on models used for analyses of relaxation function 
line shapes, fitting range of data analyses, treatment of 
single crystal versus ceramic samples, and some other factors.
The Gaussian decay, which fits most of the data from ceramic
samples quite well, is significantly different from the 
ideal field distribution $P(H)$ expected for a perfect Abrikosov 
vortex lattice in triangular lattice.  So, using a theoretical
second moment for $P(H)$ in Abrikosov lattice is not necessarily
appropriate for data analyses in real experiments.  
After various simulations and consistency checks, we 
decided to adopt a factor which gives 
$\lambda$ = 2,700 \AA\ for $\sigma$ = 1 $\mu$s$^{-1}$
for a triangular lattice.
Note that this conversion is for a standard
triangular lattice, contrary to the statements of 
Tou and collabrators \cite{touprl,touprb} who have erroneously cited that
we calculated $\lambda$ for a square vortex lattice.
Then we can derive $n_{s2d}/m^{*}$ from the observed values of
$\sigma$ and known values of $c_{int}$. 
In the horizontal axis of Fig. 7, we attach the 2-d Fermi temperature
$T_{F2d}$ corresponding to the 2-d superfluid density obtained in 
the above-mentioned procedure.

A 2-d superfluid of bose gas, such as
thin films of liquid He, undergoes superfluid to 
normal transition via  
a thermal excitation of unbound flux vortices,
as shown by Kosterlitz and Thouless (KT) \cite{ssc40}.
For paired fermion systems composed of $n$ fermions
with mass $m$, forming a superfluid with boson density $n/2$
and mass $2m$, the Kosterlitz-Thouless transition temprature
$T_{KT}$ becomes 1/8 of the 2-d Fermi temperature
$T_{F2d}$ of the corresponding fermion system.
In the KT theory, the 2-d superfluid density 
at the transition temperture $T_{KT}$ should follow
system-independent universal behavior:
namely, $(\hbar^{2}\pi)\times n_{s2d}/m^{*}$ at $T = T_{KT}$ equals $T_{F2d}/8$.
This universal relation was first confirmed by 
an experiment on He thin films \cite{KTjump}.
In systems close to ideal Bose-gas, the 2-d superfluid density
shows almost no reduction between $T=0$ and $T=T_{KT}$ \cite{KTjump}.
Thus, in such a case, we would expect the points in Fig. 7 (based on 
$n_{s2d}/m^{*}(T=0)$) to lie on 
the $T_{KT}$ line. In thin films of BCS superconductors,
the superfluid density shows much reduction from the
value of $T=0$ to $T= T_{KT}$, and the ``KT jump of
superfluid density'' becomes invisible.  
This corresponds to the situation where the points in Fig. 7 lie far in 
the right side of the $T_{KT}$ line.
In Fig. 7, most of the points lie about 
a factor of 2 away from the $T_{KT}$ line.  
The linear relation between $T_{c}$ and
$n_{s2d}/m^{*}(T=0)$ suggests relevance to the 
KT transition, as pointed out by Emery and Kivelson\cite{ssc51}.
However, the deviation from the 
$T_{KT}$ line implies serious difference from the
ideal KT situation.  

\section{MAGNETIZATION MEASUREMENTS: EXPERIMENTAL}
\label{sec:level6}

Magnetization measurements were performed using a SQUID 
magnetometer (Quantum-Design) at Columbia. 
Aligned pressed samples were sealed 
in quartz ampules. The raw response curve
was corrected by the subtraction of the quartz background curve 
measured in advance.  
In the normal state of the superconducting Hf(Zr)NCl samples, 
as well as the parent compounds, weak-ferromagnetic behavior is 
observed up to room temperature. This weak-ferromagnetic 
behavior changes by the intercalation. Therefore, we 
subtracted the weak-ferromagnetic contribution.
We estimated this by extrapolating  
temperature dependence, assuming the Curie-Weiss law 
($M = C / (T - \theta)$) 
and fitting the normal state magnetization 
in the temperature range of $2.5T_{c} \leq T \leq 5T_{c}$. 
In this temperature range, superconducting fluctuations can be 
neglected and the temperature dependence is slightly concave. 
We only used the data with the extrapolated weak-ferromagnetic 
contribution smaller than 10 \%\ of the diamagnetic magnetization, 
to avoid an error from the assumption of the Curie-Weiss 
temperature dependence. 
The model developed by Hao {\it et al.} \cite{hao} was applied 
to the analysis of the reversible region. 
In this model, reduced (dimensionless) magnetization 
$M' = M/\sqrt{2} H_{c}(T)$ and field 
$H' = H/\sqrt{2} H_{c}(T)$ scales as a single function 
that contains the Ginzburg-Landau parameter $\kappa$ 
as a unique parameter. 
In our analysis, we optimized $H_{c}(T)$, in addition to 
$\kappa$ as a parameter 
independent of temperature. 
Resistivity measurements were performed using a well-aligned 
pressed sample with four electrodes that is sealed in a cell made 
of Kapton film and epoxy glue. 

\section{MAGNETIZATION MEASUREMENTS: SUPERCONDUCTING PROPERTIES}
\label{sec:level7}

Magnetization measurements were performed in  
Li$_{0.17}$ZrNCl and Li$_{0.15}$THF$_{0.08}$ZrNCl 
with magnetic fields applied parallel to the c-axis.
Figure 8 shows the results obtained after the 
corrections for the quartz ample background and for 
the weak-ferromagnetic contribution. 
We note that a crossing point exists in the $M(T)$ curve under various 
magnetic fields for each system, which is characteristic of 
quasi-two-dimensional superconductors~\cite{kesprl}. 
There are reversible temperature regions where ZFC and 
FC magnetization curves overlap each other. Below a certain 
temperature (the pinning temperature $T_{p}$), 
ZFC and FC magnetization curves deviate. 
We notice that 
the reversible region is wider for Li$_{0.15}$THF$_{0.08}$ZrNCl. 
This result is consistent with a picture that, by the expansion of 
the interlayer distance, the interlayer coupling become weaker and 
the pinning of the vortices becomes less effective. 

In the data analyses in the reversible 
region of Li$_{0.17}$ZrNCl and Li$_{0.15}$THF$_{0.08}$ZrNCl,  
we confined to the temperature region apart from $T_{\rm c}(H)$ in 
order to avoid 
ambiguity due to the superconducting fluctuations.  
As shown in Fig. 9, the data scale quite well 
to Hao's model in the 
whole reversible temperature range below $T_{c}$ for the both 
systems.
For all the data, $M' << H'$ and hence the demagnetization 
factor can be ignored. 
This analysis yielded values of $\kappa$ ranging between 50 and 80
(see Table 1), which indicates that these  
compounds are extreme type-II superconductors. 
In Fig. 10, we show the values of the upper critical field 
$H_{c2,//c}(T)$ obtained down to $T=2$ K in this process using Hao's model. 
The temperature dependence of 
$H_{c2,//c}(T)$ fits well to an empirical formula 
$H_{c2,//c}(0) [1-(T/T_{c})^2]$, as shown by the dashed lines in Fig. 10. 
We emphasize that the low temperature
limit value $H_{c2,//c}(T\rightarrow 0)$ can be obtained almost 
without any extrapolation using this formula: the resulting
values are shown in Table 1.  The $H_{c2//c}$ value of $\sim$ 4-5 T 
in ZrNCl-Li-THF system is 
about a factor of 2 smaller than $H_{c2} \sim$ 10 T in HfNCl-Li-THF system reported
by Tou {\it et al\/} \cite{touprb}. These results might indicate that
$H_{c2}$ roughly scales with $T_{c}$.  A similar nearly linear relation between $H_{c2}$
and $T_{c}$ can be found in the $H_{c2}$ values for high-$T_{c}$ cuprate
superconductors in the optimum doping region.

The critical temperature 
$T_{c}$ obtained using Hao's model is 14.9 K for both samples, 
which agrees with the estimate from the onset of diamagnetism due to  
superconductivity. 
We notice that at $H=55$ kG above $H_{c2,//c}(0)$, 
a diamagnetic behavior due to superconducting fluctuation 
was observed in magnetization as shown in Fig. 8.
Similar results due to critical fluctuations have been reported in 
HTSC~\cite{fl1,fl2,fl3} and BEDT ~\cite{fl4} systems. 
We obtained the in-plane coherence length $\xi_{ab}(0)$ and the in-plane penetration 
depth $\lambda_{ab,M}(0)$ using expressions 
$H_{c2}(0) = \phi_0/2\pi\xi(0)^2$ and 
$\kappa = \lambda/\xi$. 
These results are also 
summarized in Table 1. We note that $H_{c2,//c}(0)$, 
and $\xi_{ab}(0)$ are almost unaffected by 
the interlayer distance.
This agrees with the view that the essence of the 
superconductivity in Hf(Zr)NCl is dominated in Hf(Zr)-N 
honeycomb double layers. 
The values of the penetration depth determined 
both from $\mu$SR and magnetization show reasonable 
agreement,  
although the former is $\sim20$ \% smaller than the latter 
for both compounds. 
 
In magentization measurements (see Fig. 8) and $\mu$SR measurements
(see Fig. 5(a)), the results
become history dependent below a pinning temperature $T_{p}$ for a given 
external field $H_{ext}$.  This feature can be expressed by defining 
the irreversibility field $H_{irr}$ for a given temperature $T$ as
$H_{ext}(T=T_{p}) = H_{irr}$.
Figure 10 also includes $H_{irr}$ as a function of 
temperature, determined from magnetization and $\mu$SR
measurements. 
The results obtained from the two different techniques
exhibit excellent agreement.   ZrNCl superconductors have a quite large 
reversible region in the $H$-$T$ plane. 
The temperature dependence of the irreversibility field  
fits well to a functional form 
$H_{irr}(T) = H_{irr}(0) (T_{c}/T-1)^n$,
obtained for three-dimensionally 
fluctuating vortices\cite{crossover},    
with $n=1.5$ at low fields below $H \sim 0.4$ Tesla. 
This provides support to our
assumption of 3-d vortex lines which
we adopted in our analyses of TF-$\mu$SR spectra taken below 
$H=0.1$ Tesla. 
At higher fields, the fitting becomes worse, similarly to 
Ref. \cite{crossover}.  This may be related to a dimensional 
crossover from 3- to 2-d vortex fluctuations. 
More careful measurements are necessary to conclude this point. 

In order to provide a cross-check for the results of $H_{c2}(T)$, 
we performed magnetoresistance measurements on  
Li$_{0.17}$ZrNCl. The temperature dependence of 
resistivity for two sets of field and current configurations 
is shown in Fig. 11. High resistivity of the order of 100 
$m\Omega cm$ and negative slope of the resistivity in the normal state at low 
temperatures could be due to grain boundaries and may not be 
intrinsic. The observed superconducting transition is broadened 
by superconducting fluctuations, weakly superconducting regions 
such as grain boundaries, and by vortex motion due to Lorenz force. 
We notice that the resistive broadening is slightly larger for 
$H // {c}$, which is a natural consequence of significant  
superconducting fluctuations only for $H // {c}$. 
We defined $T_{c}(H)$ where resisitivity shows 50 \% drop 
of the maximum value.  Figure 12 shows $H_{c2}(T)$ 
obtained in this procedure.  These absolute values of $H_{c2}(0)$ for $H//c$ 
agree reasonably well with those from magnetization measurements, 
in spite of the unreliable definition due to the broad resistive 
transition.  The difference between the temperature dependences of the 
magnetization (Fig. 10) and resistive (Fig. 12) $H_{c2}$ data 
may be due the above-mentioned limitations of the resistive measurements.
The anisotropy ratio of the upper critical field 
$H_{c2,\perp c}/H_{c2,// c}$ is roughly 3 
as shown in Fig. 12.  Although we do not have data for
the system with co-intercalation of THF or PC,
the anisotropy ratio would presumably increase in more
2-d systems with larger stacking unit distance.

\section{DISCUSSIONS AND CONCLUSIONS}
\label{sec:level8}

The quasi-two-dimensional nature of the superconducting state 
appears in various superconducting properties of 
intercalated Hf(Zr)NCl. $T_{c}$ correlates with 
a 2-d superconducting carrier density $n_{s2d}$ devided by 
effective mass $m^{*}$ rather than the 3-d counterpart. 
Diamagnetic magnetization due to superconducting 
fluctuation for $H // {c}$ is observed at high temperatures 
and high fields. The crossing point exists in $M(T)$ 
curves measured at various fields. The reversible region 
of magnetization becomes larger with the increase of 
interlayer distance, suggesting weaker interlayer coupling.  

In addition to these results, we note that $T_{c}$, 
$n_{s2d}/m^{*}$, and $\xi_{ab}$ ($H_{c2,//c}$), 
all show moderte dependence on chemical doping level 
which presumably represents the in-plane 
normal-state carrier concentration, while remaining almost independent 
of the stacking unit distance. 
These parameters are closely 
related to the superconductivity mechanism of this layered 
superconductor. Since the coherence length is a measure for 
the pair size, independence of $\xi_{ab}$ on 
interlayer distance implies that interlayer coupling does not
affect the pair formation. 
It is then possible to consider a picture
in which fluctuating superconductivity  
exists within a given layer, while 
the layers are coupled weakly by Josephson coupling to 
achieve 3-d 
bulk superconductivity. 

On the other hand, 
$\lambda_{ab}$ and $T_{p}$ are strongly affected by the 
interlayer distance. 
The reduction of $\lambda_{ab}$ with increasing stacking-unit distance
can be understood as a simple reduction of the supercurrent density
caused by lower density of the planes.
The strong dependence of $T_{p}$ on $c_{int}$ is not
surprising: this behavior has been seen in many HTSC cuprates.
 
Our results show that Hf(Zr)NCl with variable interlayer distance 
as well as carrier concentration is 
suitable for systematic studies of layered superconductors. 
In addition, low $H_{c2}$ value of this compound makes it 
easier to cover the whole 
superconducting region in the $H$-$T$ plane, and helps 
our study of vortex phase diagram. 

In Fig. 6 and Fig. 7, we have compared the results from the
nitride-chlorides with other layered superconductors.
All the arguments in the previous paragraphs, as well as 
Fig. 7, give an impression that 2-d properties are
predominantly important factor of all of these layered
superconductors.  However, comparison among
different cuprate systems having different interlayer 
spacing $c_{int}$ indicates that the 3-d 
interlayer coupling plays a very important role in 
determing $T_{c}$ in the cuprates.  Furthermore,
the observed results in Fig. 7 show about a factor 2
deviation from the $T_{KT}$ line.  These results
indicate that a simple theory for KT transition,
whose $T_{c} = T_{KT}$ is unrelated to the interlayer 
coupling, is not applicable either to the cuprates nor to 
the nitride-chloride systems.  Further experimental and 
theoretical studies are required to determine
the origin of this deviation.  Studies of crossover from Bose Einstein
to BCS condensation, in the case of 2-d systems,  
might provide a clue for understanding this feature.

The absolute values of the penetration depth $\lambda$, obtained from 
the $\mu$SR and magnetization measurements, show a reasonable agreement.
We notice, however, about 20-30 \%\ difference in the values from the 
two different methods (see Table 1).  $\mu$SR and magnetization
estimates of $\lambda$ often exhibit some disagreement of this
magnitude, as can be found also in the cases of HTSC and organic
systems.  Ambiguity of $\lambda$ with 20-30 \%\ would, however,  
correspond to $\sim$ 50 \%\ ambiguity in the estimate of the 
superfluid density.  In this situation, it would be ideal 
if there were a method to cross-check the superfluid density
derived from $\mu$SR results in a completely different perspective.  

In Bi2212 cuprate systems, Corson {\it et al.\/} \cite{ssc53}
measured the frequency dependent superfluid response,
and found temperature $T_{KT}$ above which the superfluid density
depends on the measuring frequency.  The superfluid density observed at
$T=T_{KT}$ was consistent with the value expected in the universal
argument of KT.  This provides an excellent system-independent calibration 
to the superfluid density.  The difference between the superfluid density
at $T_{KT}$ and at $T\rightarrow 0$ should correspond to the 
distance (in the horizontal direction) 
of the corresponding data point in Fig. 7 from the $T_{KT}$ line.
The $\mu$SR Bi2212 data point in a plot like Fig. 7 lie about a factor
2-3 away from the $T_{KT}$ line.  This factor agrees reasonably well 
with the reduction of the superfluid density from the $T=0$ value
to the $T=T_{KT}$ value observed by Corson {\it et al.\/} \cite{ssc53} in Bi2212
system in a similar doping region.  This satiafactory cross-check for the 
Bi2212 system indicates that our choice of the conversion factor
between $\sigma$ and $\lambda$ was reasonable, and the superfluid
density derived by $\mu$SR is very reliable.
Of course, comparisons among $\mu$SR data for different
systems in a relative scale can be
performed without being affected by an ambiguity of their absolute
values of the superfluid density.

We performed $\mu$SR measurements on three 
different specimens based on ZrNCl with Li concentraitons 0.15, 0.17 and 0.4, and found 
that the results of 2-d superfluid density $n_{s2d}/m^{*}$ for these systems do not 
show much difference among one another.  This phenomenon could be explained by two 
different possibilities: (a) not all the Li atoms donnate carriers on the ZrN
planes, and the Li concentration $x$ does not serve as an indicator
of normal-state carrier concentraion; 
or (b) all the Li atoms donnate electrons to the ZrN planes, but
only a finite fraction of those normal-state carriers participate in the superfluid.
The situation (b) is similar to the case of overdoped HTSC cuprates~\cite{sscreview,uemuraodssc},
where an energy-balance in the condensation process seems to determine the 
superfluid density.  Further experiments on normal-state transport properties
are required to distinguish between (a) and (b) in the nitride-chloride systems. 

In conclusion, we have synthesized and characterized several different
specimens of intercalated nitride-chloride superconductors,
and performed $\mu$SR and magnetization measurements.
The superconducting transition temperature $T_{c}$ and 
the upper critical field $H_{c2,//}$ exhibit a 
nearly linear relation with the 2-d superfluid density
$n_{s2d}/m^{*}$, while showing 
almost no dependence on the stacking unit distance.
These features suggest a highly two-dimensional nature
of superconductivity in the nitride-chloride system.

\section{ACKNOWLEDGEMENT}
\label{sec:level9}

We are grateful to A.R.~Moodenbaugh for x-ray rocking curve 
measurement, Y.~Mawatari for discussion about reversible 
magnetization, and H. Stormer for help in resistance
measurements.
This work was supported primarily by the 
National Science and Engineering Initiative of the 
National Science Foundation under NSF Award CHE-01-17752.
The work at Columbia was also supported by NSF-DMR-01-02752 and
NSF-INT-03-14058. The work at Hiroshima Univ. was supported by the
Grant-in-Aid for Scientific Research (B) (No.14350461) and the COE Research
(No. 13E2002)) of the Ministry of Education, Science, Sports, and Culture of
Japan, and by 
CREST, Japan Society for Science and Technology (JST).
Research at McMaster is supported by NSERC and CIAR (Quantum Materials Program).

\newpage

\begin{figure}

\begin{center}

\mbox{\psfig{figure=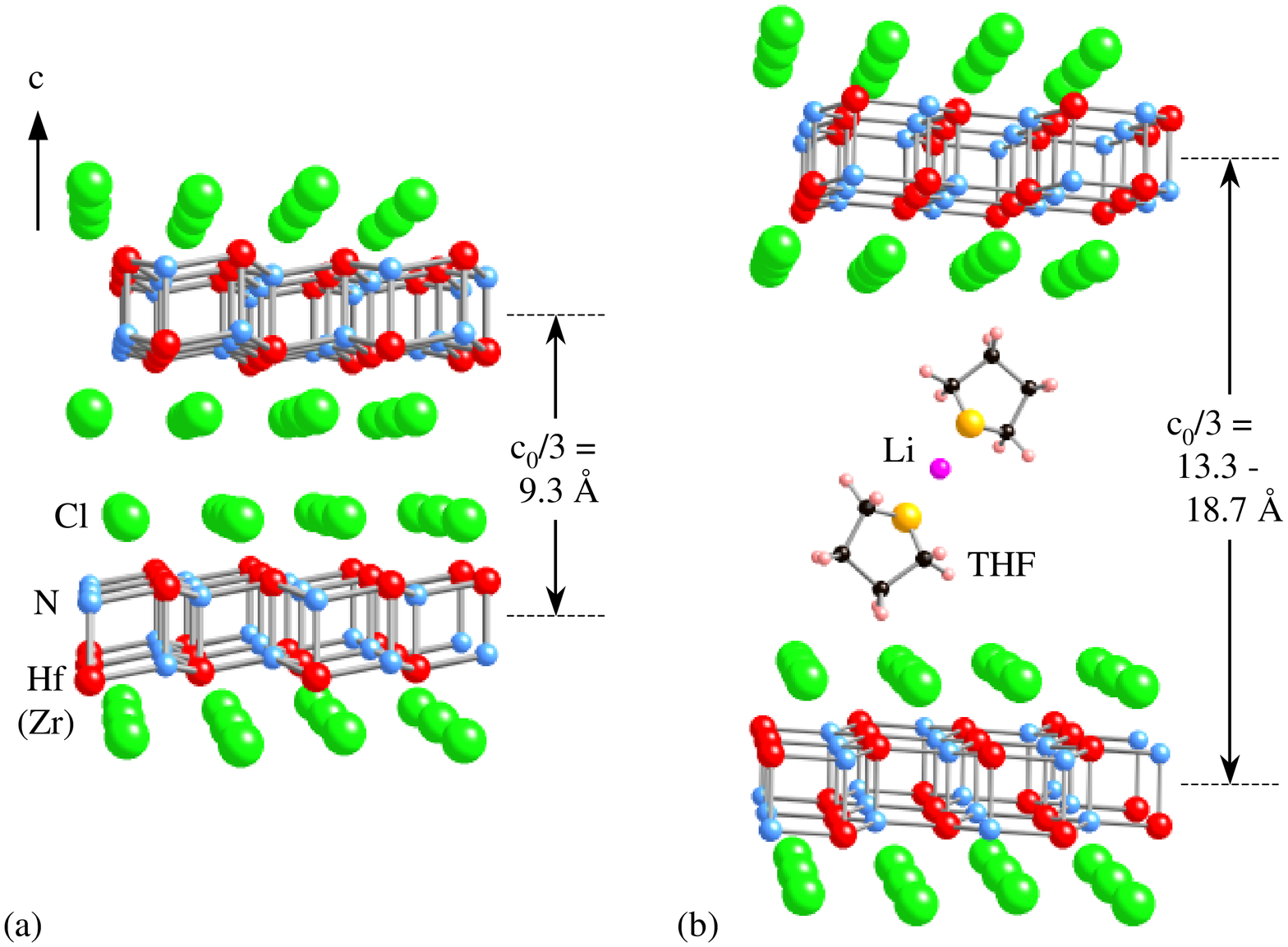,angle=270,width=6in}}

\vskip 1.0 truein

\label{Figure 1.} 

\caption{Schematic figures of the crystal structure of 
Hf(Zr)NCl (a) without intercalation and (b) co-intercalated with Li and 
THF. The stacking unit thickness ($=c_0/3$, where $c_0$ is c-axis 
lattice constant) is also shown. }

\end{center}

\end{figure}

\newpage

\begin{figure}

\begin{center}

\mbox{\psfig{figure=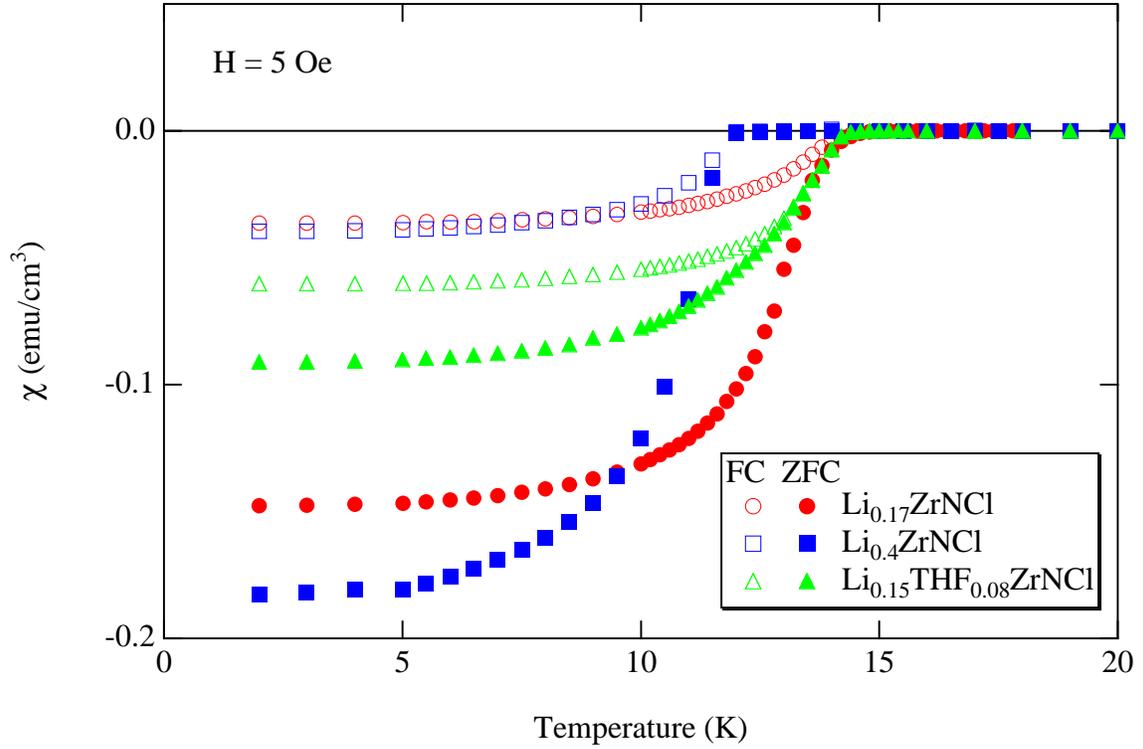,width=6in}}

\vskip 1.0 truein

\label{Figure 2.} 

\caption{Magnetic susceptibility $\chi$ vs.\ temperature 
measured in $H=5$ Oe in the field-cooling (FC) and zero-field-cooling
(ZFC) procedures in  
the specimens
of intercalated ZrNCl which were used in the $\mu$SR measurements.
}

\end{center}

\end{figure}

\newpage

\begin{figure}

\begin{center}

\mbox{\psfig{figure=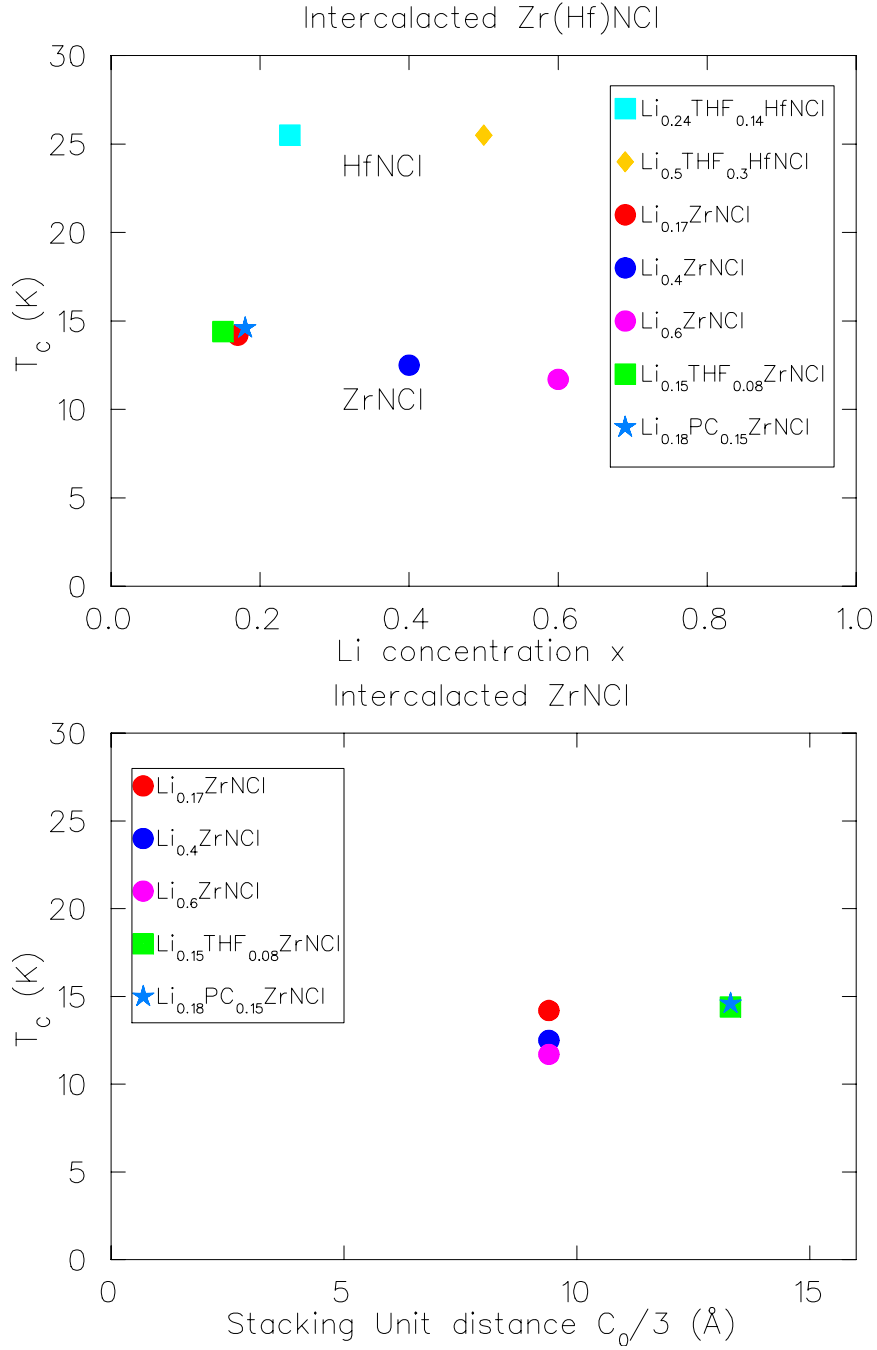,width=6in}}

\vskip 0.3 truein

\label{Figure 3.} 

\caption{Dependence of $T_{c}$,
as determined from the susecptibility, 
on (a) Li concentration $x$, and on (b) 
stacking-unit-distance $c_{o}/3$ 
in Li$_{x}$ZrNCl and Li$_{x}$HfNCl with/without
co-intercalation of THF or PC.}

\end{center}

\end{figure}

\newpage

\begin{figure}

\begin{center}

\mbox{\psfig{figure=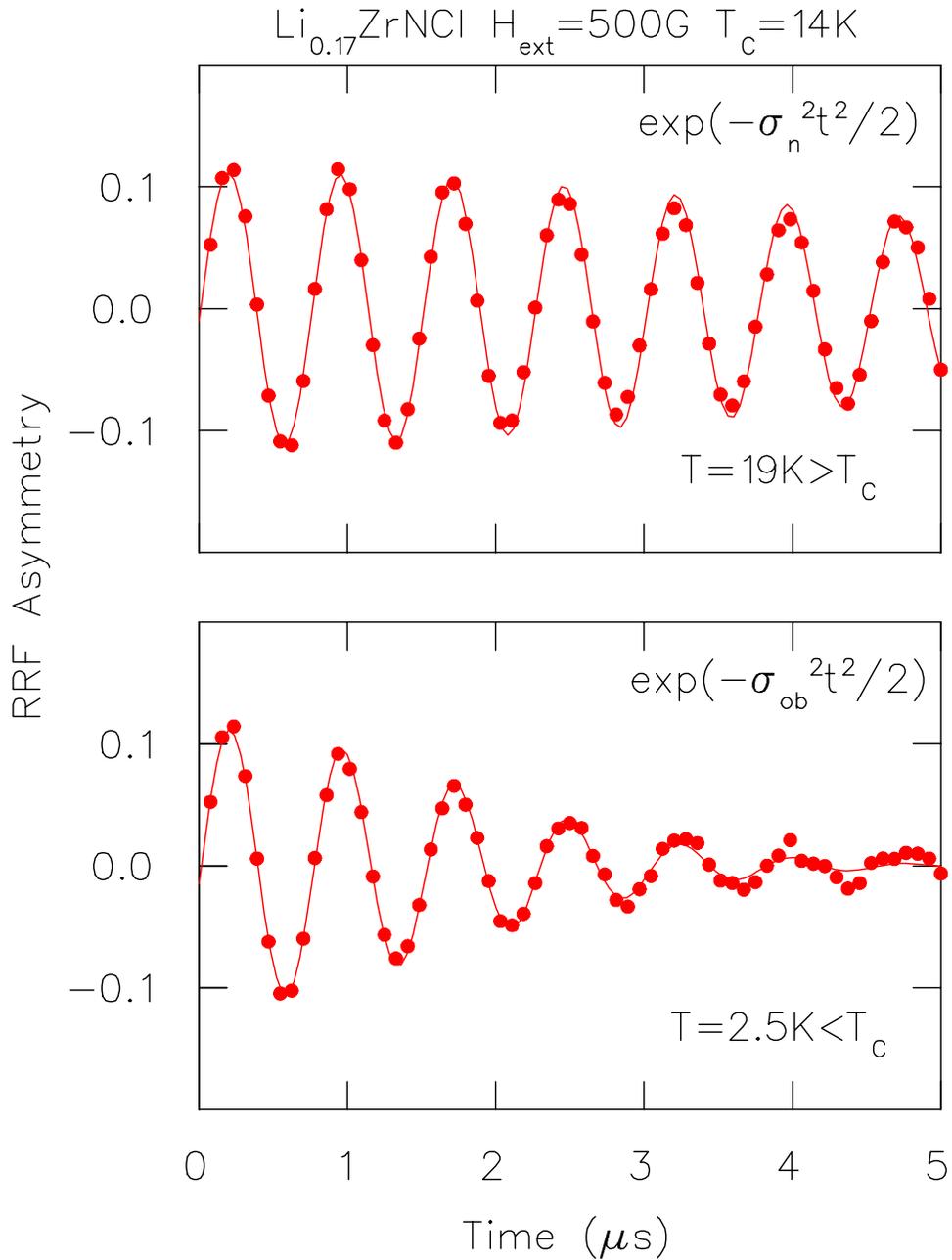,width=5in}}

\vskip 0.5 truein

\label{Figure 4.} 

\caption{The time spectra of muon asymmetry $A(t)$ 
measured in Li$_{0.17}$ZrNCl 
at (a) $T=19$ K (above $T_{c}$) and (b) $T=2.5$ K (well below $T_{c}$) 
under the transverse external field of 500 G.  
The apparent precession frequency in this graph is modified from the 
actual precession frequency by the use of a rotating reference frame.  
The solid lines 
show a fit to Gaussian decay envelope.}

\end{center}

\end{figure}

\newpage

\begin{figure}

\begin{center}

\mbox{\psfig{figure=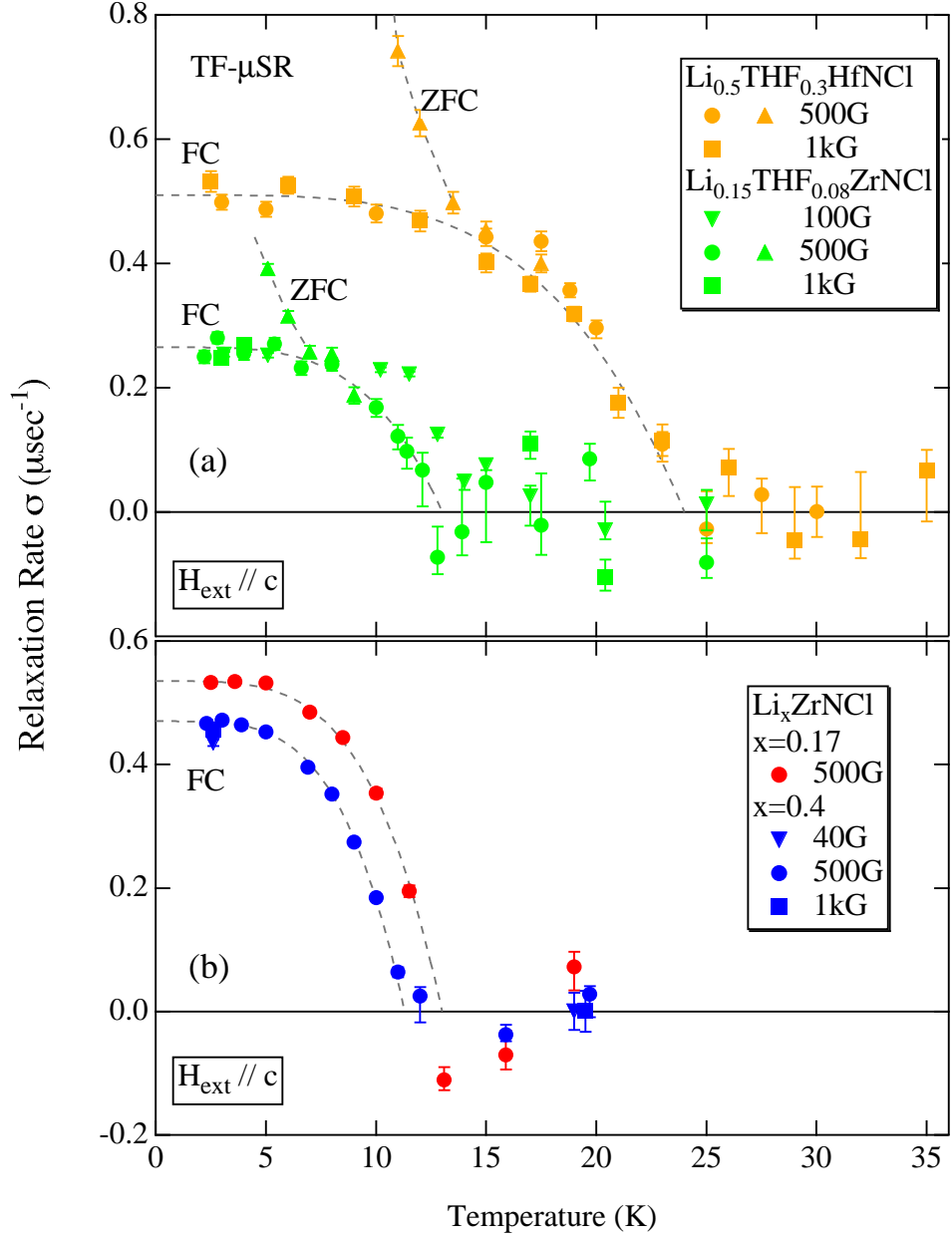,width=5in}}

\vskip 1.0 truein

\label{Figure 5.} 

\caption{Temperature dependence of the muon spin relaxation 
rate $\sigma(T)$ in (a) Hf(Zr)NCl co-intercalated with Li atoms and 
THF,
and (b) Li$_{x}$ZrNCl without co-intercalation. 
The upper panel (a) shows the results obtained in 
the field cooling (FC) and zero field cooling 
(ZFC) procedures, while the results in the lower panel
(b) were obtained in the
FC procedure. 
Dashed lines are guides to the eyes.}

\end{center}

\end{figure}

\newpage

\begin{figure}

\begin{center}

\mbox{\psfig{figure=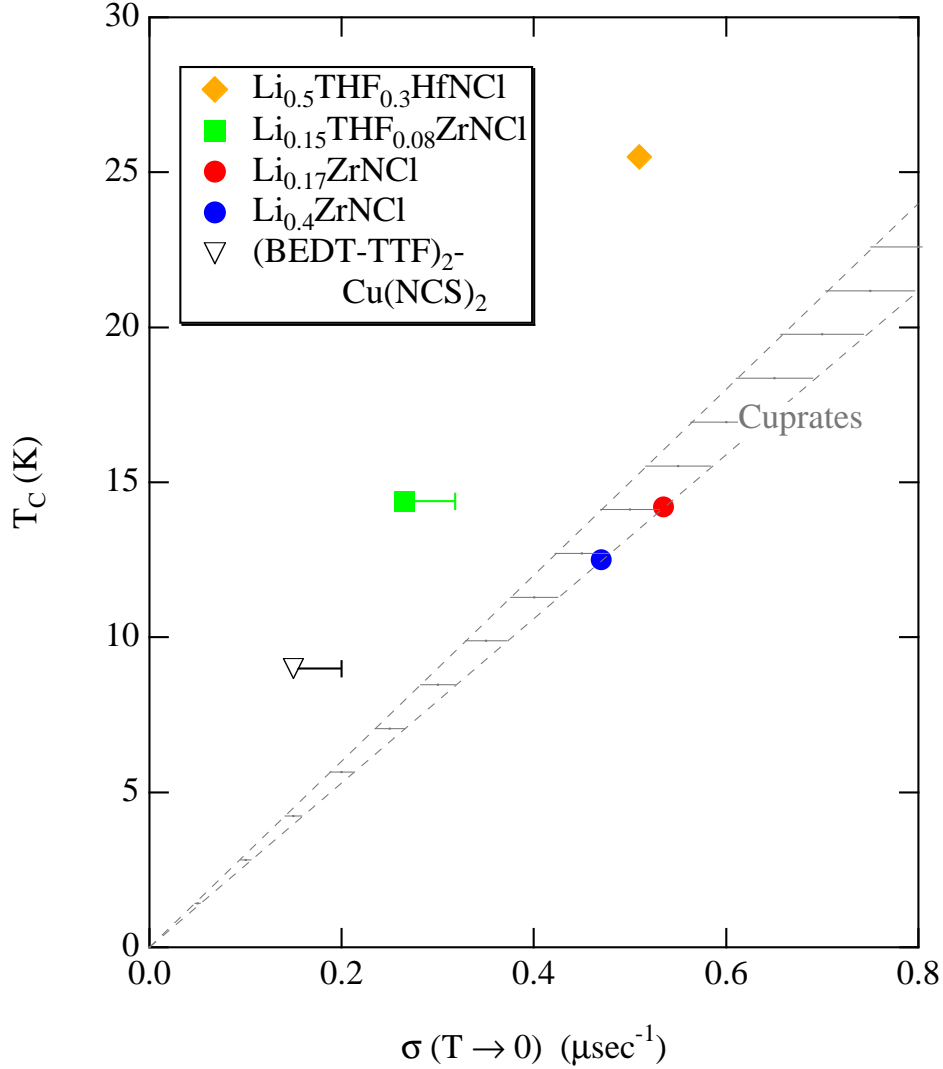,width=5in}}

\vskip 1.0 truein

\label{Figure 6.} 

\caption{Correlations between 
$T_{c}$ and the muon spin relaxation
rate $\sigma(T\rightarrow 0)$ 
of intercalated Hf(Zr)NCl (present work), high-$T_{c}$ 
cuprates~\protect\cite{ssc6}  and
(BEDT-TTF)$_{2}$-Cu(NCS)$_{2}$~\protect\cite{ssc19}.
The horizontal axis is
proportional to the 3-dimensional superfluid
density $n_{s}/m^{*}$ in the ground state. 
The results of $\sigma$ 
for the cuprates, obtained using un-oriented ceramic
specimens, are multiplied by a factor $\sim$ 1.4 for the 
comparison with those from nitride-chloride and BEDT systems
obtained using single crystals and oriented ceramic 
specimens.}

\end{center}

\end{figure}

\newpage

\begin{figure}

\begin{center}

\mbox{\psfig{figure=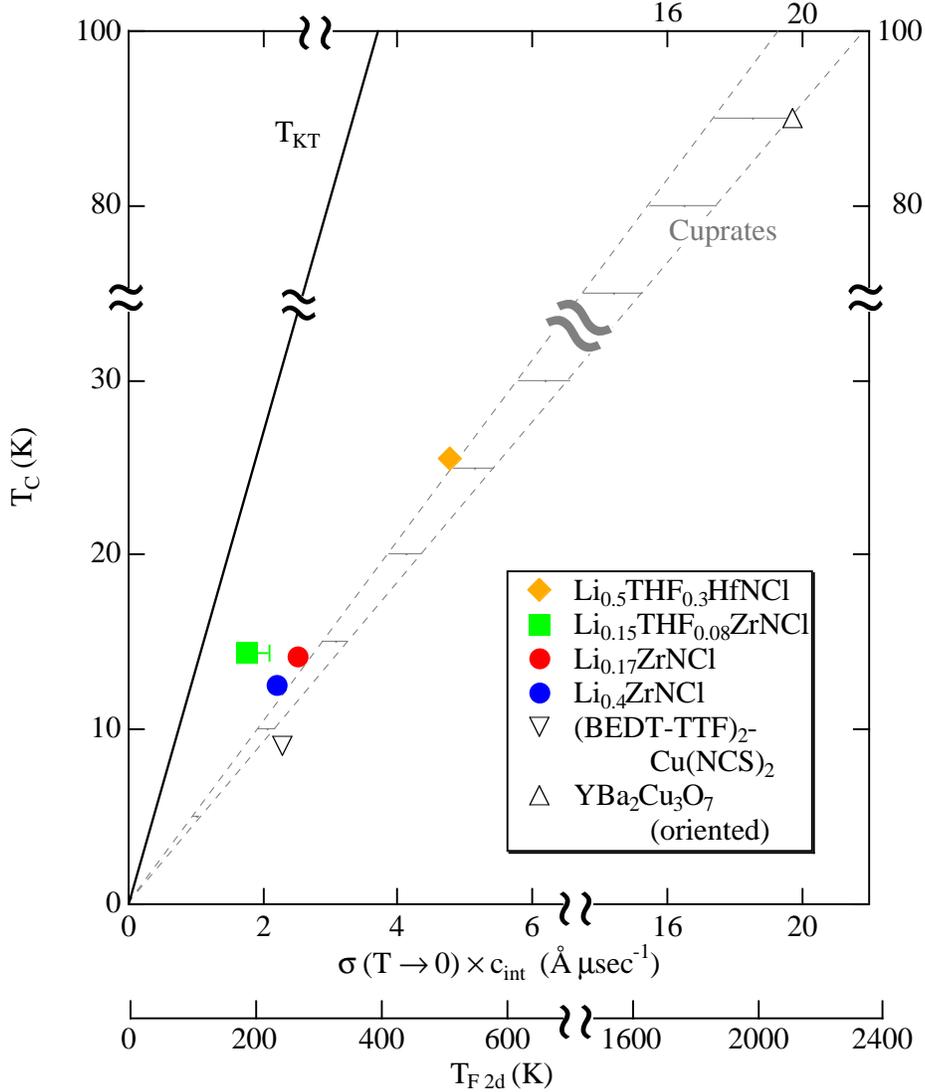,width=5.0in}}

\vskip 0.5truein

\label{Figure 7.} 

\caption{Correlations between $T_{c}$ and 
$\sigma(T\rightarrow 0) \times c_{int}$  
of intercalated Hf(Zr)NCl,
organic BEDT \protect\cite{ssc19} and YBa$_{2}$Cu$_{3}$O$_{7}$ 
(YBCO) \protect\cite{YBC_oriented}, 
where $c_{int}$ stands for average interlayer distance. 
We regard 
the double layers in the nitride-chlorides and cuprates
as two separate layers, and thus assume
$c_{int} = c_{o}/6$ for the nitride chlorides and 
$c_{int} \sim 6$\AA\  for YBCO.
The horizontal axis is proportional to the 2-d superfluid
density $n_{s2d}/m^{*}$.  To the horizontal axis, we also 
attach the corresponding 
energy scale, 2-dimensional Fermi temperature $T_{F2d}$,
obtained from the 2-d superfluid density.}

\end{center}

\end{figure}

\newpage

\begin{figure}

\begin{center}

\mbox{\psfig{figure=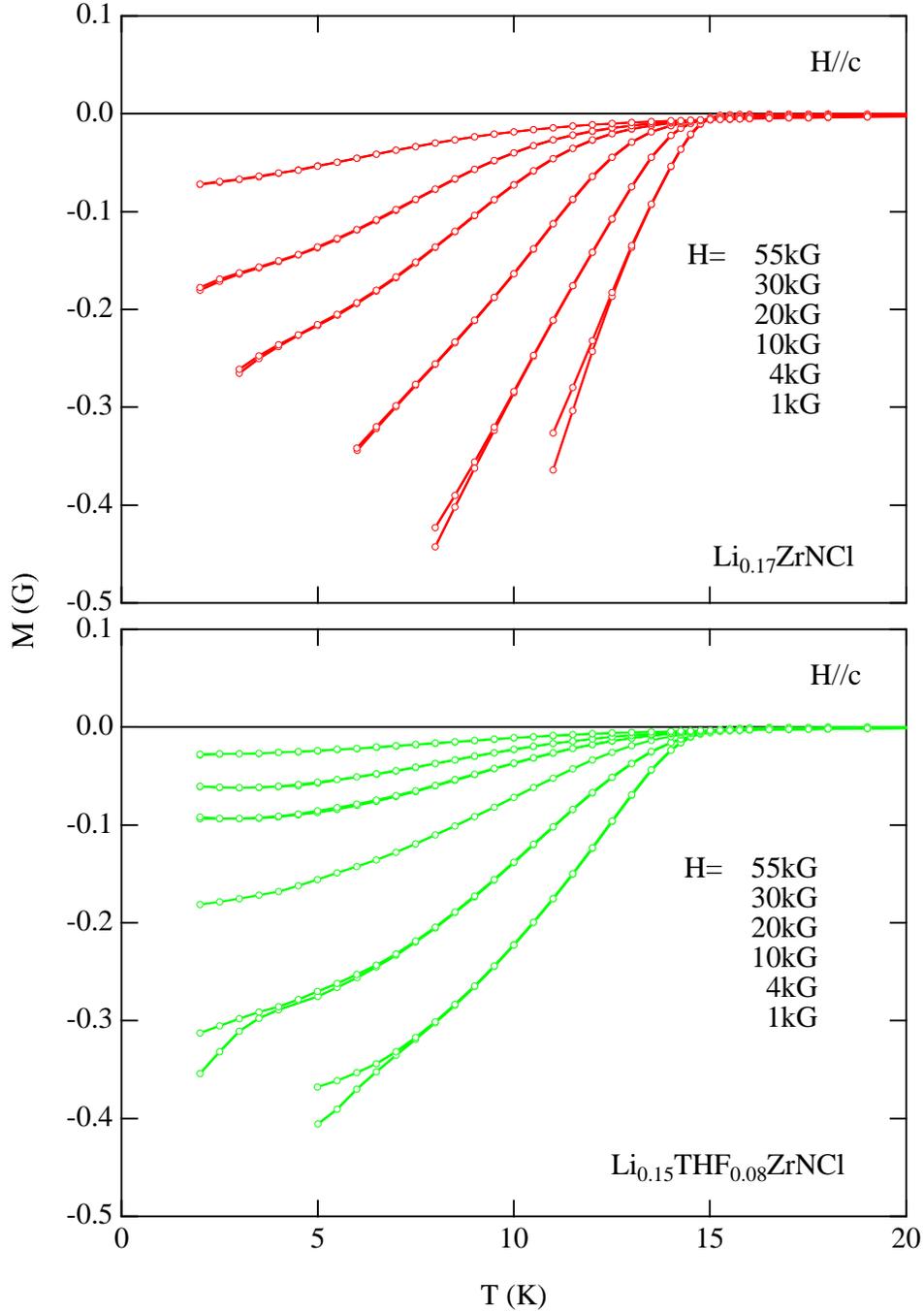,width=5in}}

\vskip 0.3 truein

\label{Figure 8.} 

\caption{Temperature dependence of the magnetization of 
(a) Li$_{0.17}$ZrNCl and (b) Li$_{0.15}$THF$_{0.08}$ZrNCl, 
in external magnetic fields applied parallel to the c axis. 
For the plotted data, the quartz ampule background is corrected 
and the weak-ferromagnetic contribution is subtracted. 
}

\end{center}

\end{figure}

\newpage

\begin{figure}

\begin{center}

\mbox{\psfig{figure=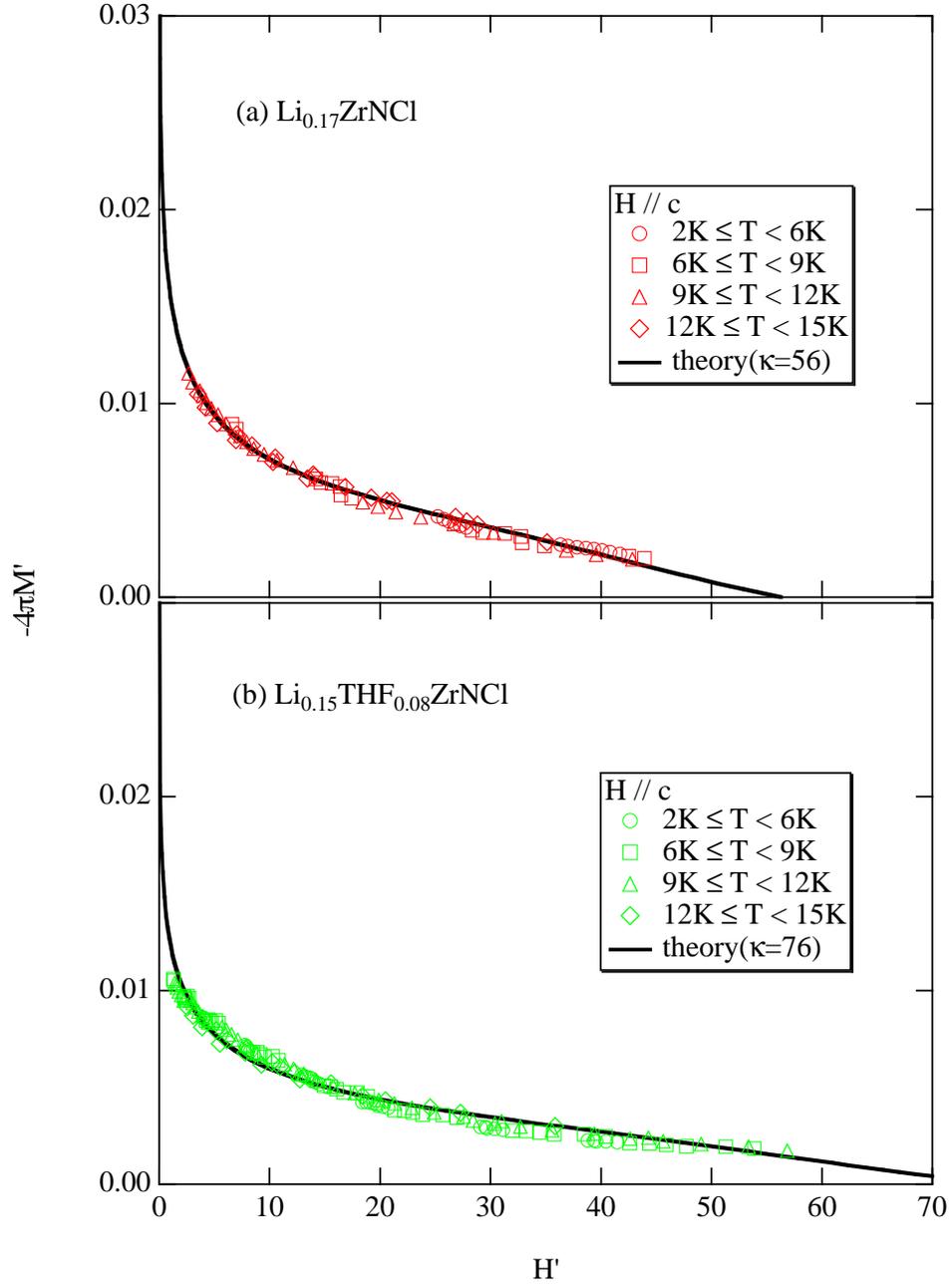,width=5in}}

\vskip 1.0 truein

\label{Figure 9.} 

\caption{Magnetization as a function of applied field (parallel to the 
c-axis) in 
(a) Li$_{0.17}$ZrNCl and (b) Li$_{0.15}$THF$_{0.08}$ZrNCl, 
shown in the reduced (dimensionless) units. The solid curves
represent a fit to the model of Hao {\it et al.\/} \protect\cite{hao}}

\end{center}

\end{figure}

\newpage

\begin{figure}

\begin{center}

\mbox{\psfig{figure=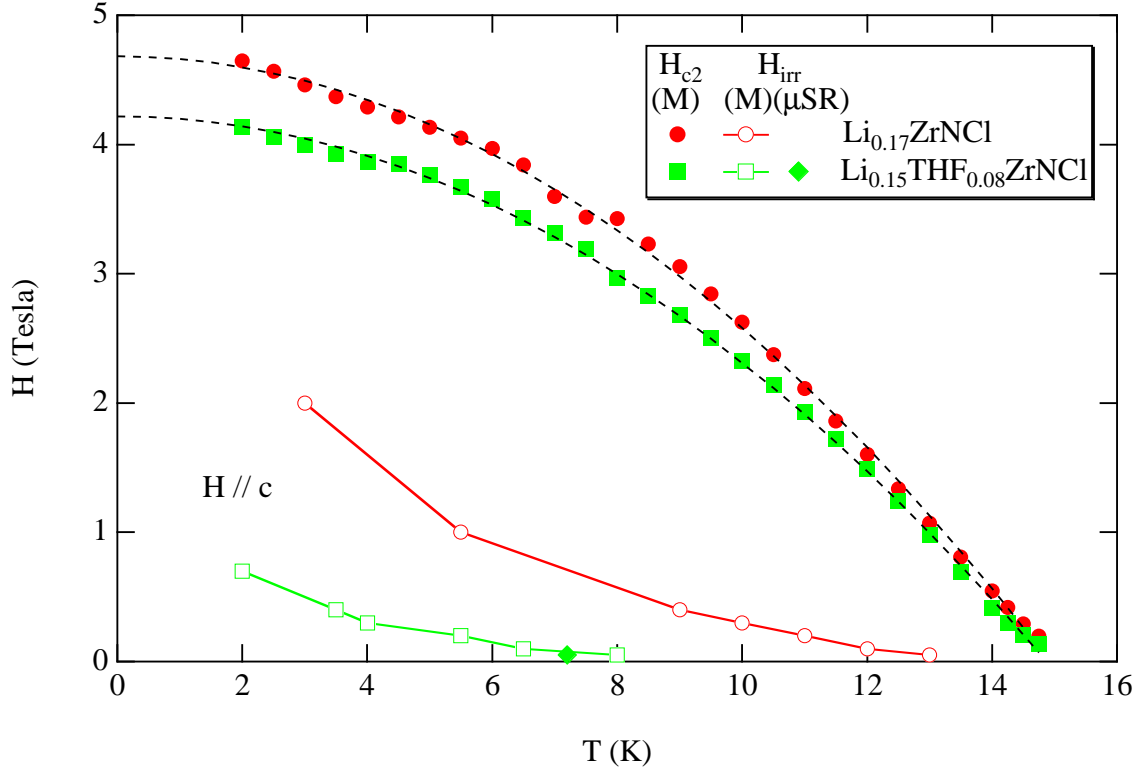,width=6in}}

\vskip 1.0 truein

\label{Figure 10.} 

\caption{Temperature dependence of the upper critical field 
$H_{c2,//c}$ and the irreversibility field $H_{irr}$ in   
Li$_{0.17}$ZrNCl and Li$_{0.15}$THF$_{0.08}$ZrNCl. 
$H_{c2,//c}$ was obtained from magnetization measurements 
(abbreviated as M in the figure), and $H_{irr}$ 
from magnetization and TF-$\mu$SR measurements.
The broken lines show a fit to the functional form of $H_{c2}$
given in the text. }

\end{center}

\end{figure}

\newpage

\begin{figure}

\begin{center}

\mbox{\psfig{figure=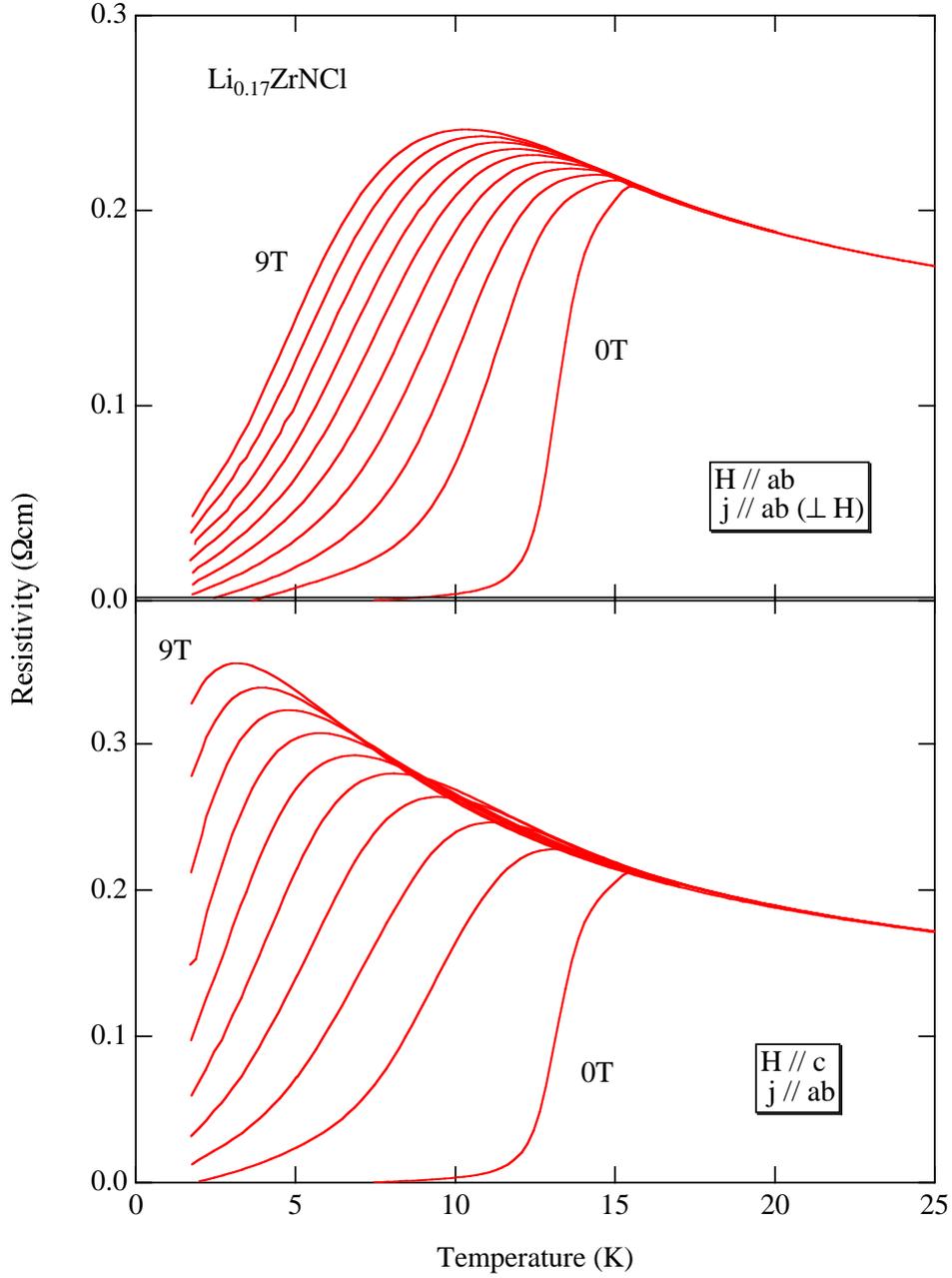,width=5in}}

\vskip 1.0 truein

\label{Figure 11.} 

\caption{Temperature dependence of the resistivity of 
Li$_{0.17}$ZrNCl under external magnetic field for two sets of the  
field and current directions, meaured using a four-probe contact.
The applied fields range from 0 to 9 T, in the intervals of 1 T. }

\end{center}

\end{figure}
\newpage

\begin{figure}

\begin{center}

\mbox{\psfig{figure=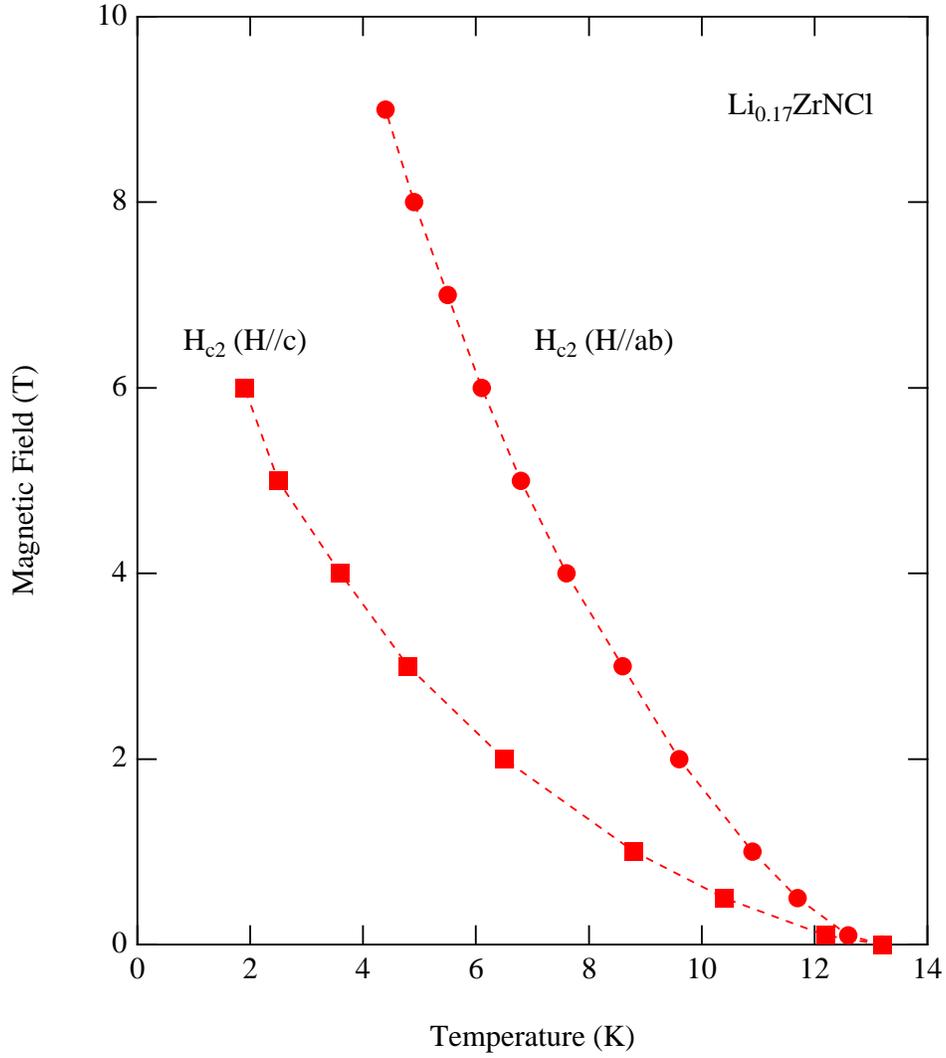,width=5in}}

\vskip 1.0 truein

\label{Figure 12.} 

\caption{Temperature dependence of the upper critical field 
$H_{c2}$ of Li$_{0.17}$ZrNCl, obtained from 
magnetoresistance. }

\end{center}

\end{figure}

\newpage

\begin{table}
\caption{Synthesis method (see the text for details), 
stacking unit distance ($c_{0}/3$), 
superconducting transition temperature $T_{c}$
estimated from magnetization, magnetic field penetration depth
$\lambda_{ab,\mu{SR}} (T \to 0)$ at zero temperature limit estimated 
from TF-$\mu$SR, 
Ginzburg-Landau parameter $\kappa$, 
upper critical field at zero temperature $H_{c2,//c}(0)$, 
coherence length at zero temperature $\xi_{ab}(0)$, and 
magnetic penetration depth at zero temperature 
$\lambda_{ab,M}(0)$ estimated from reversible magnetization 
for intercalated HfNCl and ZrNCl systems reported in the present work. 
TF-$\mu$SR and magnetization measurements were performed under 
the magnetic field parallel to the $c$-axis.}
\vskip 1.0truecm
\label{table1}
\begin{tabular}{llccccccc}
{Sample}&{Synthesis method}&{$c_0/3$}&
{$T_{\rm c}$}&{$\lambda_{{\rm ab,}\mu{\rm SR}}  (T \to 0)$}&
{$\kappa$}&{$H_{c2,//c}(0)$}&{$\xi_{\rm ab}(0)$}&
{$\lambda_{\rm ab,M} (0)$}\\
{}&{}&{\AA}&{K}&{\AA}&{}&{T}&{\AA}&{\AA}\\
% \shortstack{Synthesis\\method}
\tableline
Li$_{0.17}$ZrNCl&{(i) $n$-Butyllithium}&9.4&14.2&3700&
56&4.7&83&4700\\
Li$_{0.4}$ZrNCl&{(i) $sec$-Butyllithium}&9.4&12.5&3900\\
Li$_{0.6}$ZrNCl&{(i) $tert$-Butyllithium}&9.4&11.7\\
%#5
Li$_{0.15}$THF$_{0.08}$ZrNCl&{(ii) THF}&13.3&14.4&5200&
76&4.2&88&6700\\
Li$_{0.18}$PC$_{0.15}$ZrNCl&{(ii) PC}&13.3&14.6&\\
Li$_{0.24}$THF$_{0.14}$HfNCl&{(iii) 8 $mM$ Li-Naph}&
13.3&25.5\\
Li$_{0.5}$THF$_{0.3}$HfNCl&{(iii) 100 $mM$ Li-Naph}&
18.7&25.5&3900\\
\end{tabular}
\end{table}
\end{document}